\documentclass[aps,prd  ,twocolumn,flushleft,amsmath,amssymb,nofootinbib]{revtex4}

\usepackage{graphicx}
\usepackage{dcolumn} 
\usepackage{bm}      

\newcommand{\bq}{\begin{equation}}
\newcommand{\eq}{\end{equation}} 
\newcommand{\ba}{\begin{eqnarray}}
\newcommand{\ea}{\end{eqnarray}}

\def\kms{{\rm km s}^{-1}}

\def\Mpc{{\rm Mpc}}

\def\GeV{{\rm GeV}}

\begin{document}
\title{Cosmological Constraints on Invisible Decay of Dark Matter} 
\author{Yan Gong}
\author{Xuelei Chen}
\affiliation{National Astronomical Observatories, Chinese Academy of
  Sciences, 20A Datun Rd, Chaoyang District, Beijing 100012, China}

\begin{abstract}
The cold dark matter may be in a meta-stable state and decays to other
particles with a very long lifetime. 
If the decaying products of the dark matter are weakly
interacting, e.g. neutrinos, then it would have little impact on 
astrophysical processes and is therefore difficult to observe.
However, such a decay would affect the expansion history of the Universe
because of the change of the equation of state. 
We utilize a high-quality type Ia supernovae (SN Ia) data set 
selected from several resent observations and the position of 
the first peak of the Cosmic Microwave Background (CMB) 
angular spectrum given by the WMAP three-year data
to constrain the dark matter decay-to-neutrino 
rate $\Gamma=\alpha \Gamma_{\chi}$, where $\alpha$ is the fraction of the
rest mass which gets converted to neutrinos, and $\Gamma_{\chi}$ is the
decay width. We find that $\Gamma^{-1} > 0.7\times10^3$ Gyr at 
95.5\% confidence level.

\end{abstract}

\maketitle

\section{Introduction}

The nature of dark matter is a great mystery of the Universe. 
Although it constitutes about a quarter of the total cosmic density, 
its many properties are still not known. 
Various kinds of dark matter candidates have been proposed, 
for instance, the weakly interacting massive particle (WIMP) is one of
the most popular candidates \cite{dmreports}.
However, the standard Cold Dark Matter (CDM) still have 
some problems at small scales, namely
the halo cuspy profile problem and the missing satellite 
problem \cite{NFW96,M98}, which motivated researchers to study 
other possibilities \cite{OS03}. 

Decaying cold dark matter model (DCDM) is one of such possibility. 
It has been suggested that DCDM may help resolve the small scale 
problems of the standard CDM \cite{C01}. Furthermore, from the particle physics
perspective, it is also natural
 to think that the dark matter particle may be a meta-stable
particle, which could decay into more stable particles. Such
possibility could be realized if the dark matter is not the true
Lightest Supersymmetric Particle (LSP) \cite{axino}, or if the R-parity of the 
LSP is not exactly conserved \cite{R-break}. 

If the decaying products are photons, electrons, quarks or 
other particles which are involved in the strong or electromagnetic 
interactions, then it is relatively easy to be observed. Stringent
limit on radiative particle decay have been derived, e.g., from the
reionization history of the Universe \cite{CK04,zhang07,reiondecay}.  However, if
the decaying products include only weakly interacting particles, e.g. 
neutrinos or other WIMPs, then it is more difficult to obtain such 
constraint.

We consider a generic unstable cold dark matter particle which decays 
invisibly, producing only weakly interacting particles. Within the
standard model of particle physics, 
the only such weakly interacting particle is
neutrino. The unstable decaying particle may also produce another 
weakly interacting particle which is beyond the standard model. For
example, if the sneutrino is a meta-stable particle, it may decay to 
a neutralino and a neutrino, conserving both lepton number and R-parity.

Since neutrinos have very small masses, those produced in such decays will be
highly relativistic. We assume here that the recoil on the heavier 
decaying product  is small so that it could still be treated as part of CDM.
As the non-relativistic DCDM decays to relativistic 
neutrinos, the equation of state changes 
from 0 to 1/3, and this will affect the expansion rate $H$. 
Observational constraint on relativistic component were discussed 
previously, see e.g. Ref.~\cite{ZW01,PC04,IOT04}.

Here, we constrain the invisible decay of dark matter using two
observations which probe the cosmic expansion rate. 
The observations we use include (1)
type Ia supernovae (SN Ia) data selected from several resent 
observations (involving Gold06 \cite{Riess06}, SNLS \cite{Astier05} and 
ESSENCE \cite{Wood07}); (2) the position of the first peak of 
the Cosmic Microwave Background (CMB) angular spectrum $\ell_1$
given by the three-year WMAP observations \cite{Hinshaw06,Spergel06}. 
These two observations cover a large redshift range ($z_{\rm SN_{sel}}\to1.7$
and $z_{\rm CMB}\to1048$) and could give a tight and reliable constraint
on the DCDM. Throughout the paper, we assume the dark energy is a cosmological constant 
and the geometry of the Universe is flat.

\section{Constraints}
The evolution equations for the energy density of 
DCDM particle $\rho_{\chi}$ and its decay products 
$\rho_{\nu}$ are given by
\begin{eqnarray} 
\dot{\rho}_{\chi} + 3H\rho_{\chi}  &=&  -\Gamma \rho_{\chi},\\
\dot{\rho}_r + 4H\rho_r  &=&  \Gamma \rho_{\chi}, 
\end{eqnarray}
where $\Gamma=\alpha \Gamma_{\chi}$, $\Gamma_{\chi}$ is the decay width of
the particle, $0<\alpha<1$ is the fraction of the rest mass of the dark matter
particle which goes to relativistic neutrinos during each decay.  
The Hubble expansion rate $H$ is given by 
\bq \label{eq:h} H^2 = \frac{8\pi G}{3} (\rho_{\chi}+\rho_b+\rho_r+\rho_{\Lambda}), \eq
here $\rho_b$ is the baryon energy density which evolves as $(1+z)^3$,
$\rho_r$ is the relativistic particle energy density, 
which in the absence of decaying contribution would evolves as $(1+z)^4$,
and $\rho_{\Lambda}$ is the energy density of cosmological constant. 
Substitute $\frac{d}{dt} = -H(1+z)\frac{d}{dz}$ into Eq.~(\ref{eq:h}), 
the redshift evolution of the energy density for different components
can be solved numerically.
We denote $\Omega_{i0}=\rho_{i0}/\rho_{c0}$, where $\rho_{c0}$ is the
critical density at $z=0$, and take the radiation background (photon
and neutrinos produced in the early Universe) in the absence of 
decaying contribution as $\hat{\Omega}_{r0}=4.183
\times 10^{-5} h_0^{-2}$ where $h_0=H_0/100~\kms \Mpc^{-1}$ 
is the reduced Hubble constant. The baryon density is given by 
the 
CMB \cite{Hinshaw06,Spergel06} and big bang nucleosynthesis 
measurements \cite{Kirkman03}, though the actual value does not 
significantly affect our result.
The cosmological model is then described by the parameter set 
\bq
\theta = (\Omega_{\chi 0},~\Omega_{\nu 0},~\Gamma,~h_0)\nonumber.
\eq

The type Ia supernovae can be used as ``standard candles''. We have
selected a sample of 182 high quality SN Ia data 
from several resent observations, 
including 30 HST supernovae, 47 SNLS supernovae from the Gold06 data set, 
60 ESSENCE supernovae, and 45 nearby supernovae from WV07 \cite{Nesseris06}.
The MLCS2k2 light-curve fitter \cite{Riess96,mlcs2k2} is used to
process this data set. The $\chi^2$ for SN Ia data is
\bq
\label{eq:chisq} \chi^2_{\rm SN_{sel}}({\bf \theta}) = 
\sum_{i=1}^{N}\frac{(\mu_{obs}(z_i)-\mu_{th}(z_i))^2}{\sigma_i^2},
\eq
where $\mu_{obs}$ and $\sigma_i$ are the observational modulus and its error, 
and $\mu_{th}$ is the theoretical modulus given by
\bq
\mu_{th}(z) = 5\log_{10}d_L(z) + 25,              
\eq
where the luminosity distance can be computed as
\bq
 d_L({z;\bf \theta}) = (1+z)\int_{0}^{z}\frac{cdz'}{H(z')}.
\eq

In order to constrain the evolution at high redshift, 
we also use the position of the first peak of the 
CMB angular power spectrum 
$\ell_1$ as measured by the WMAP three-year data.
An alternative way of using CMB to constrain cosmic 
expansion history without invoking a full
CMB calculation is to use the so called
shift parameter $R$ \cite{Wang06}, which is a geometric test which assumes
that the size of the sound horizon at recombination is 
fixed \cite{Elgaroy07, Carneiro07}. However, the shift parameter 
was derived for assuming constant CDM density, which might not 
apply in our case \cite{Efstathiou99}. 

The angular scale of the sound horizon at last scattering,
$\ell_A\equiv\pi r(z_{ls})/r_s(z_{ls})$, is the ratio of 
the comoving angular diameter distance of the last scattering surface 
$r(z_{ls})$ to the comoving size of the acoustic horizon $r_s(z_{ls})$ 
at decoupling, which denotes both the size of sound
horizon and the geometrical property of the Universe. 
Relating to the cosmic expansion rate, 
this acoustic scale can be written as \cite{Page03}
\bq
\ell_A = \frac{\pi \int_0^{z_{ls}}\frac{cdz'}{H(z')}}{\int_0^{1/(1+z_{ls})}\frac{c_sda'}{H(a')a'^2}}
\eq
where $z_{ls}$ is the redshift of last scattering, $a$ is the scale factor 
and
\bq
c_s = c\Big(3+\frac{9}{4}\frac{\Omega_{b0}a}{\Omega_{\gamma0}}\Big)^{-1/2}
\eq
is the sound speed \cite{Carneiro07}. Here, $\Omega_{\gamma0}=\Omega_{r0}/1.681$ is 
the energy density of photon today. The value of $\ell_A$ is related 
to $\ell_1$ \cite{Page03}, the first peak in the CMB angular power spectrum.
Following Hu et al. \cite{Hu01}, we use the relation
\bq
\ell_1 = \ell_A(1-\phi_1),
\eq
where $\phi_1$ is a phase factor, and if we set the spectral index
$n=1$ and $\Omega_{b0}h_0^2=0.02$, then it is given by the fitting
formula:
\bq
\phi_1 = 0.267\big(\frac{r_{\ast}}{0.3}\big)^{0.1}.
\eq
Here $r_{\ast}\equiv\rho_{\gamma}(z_{ls})/\rho_m(z_{ls})$ and
$\rho_m(z_{ls})=\rho_b(z_{ls})+\rho_{\chi}(z_{ls})$.
This fitting formula is consistent with the other forms 
\cite{Doran02} and is a good approximation for our model, 
especially for the long lifetime DCDM. 

According to the measurement of the three-year WMAP data 
\cite{Hinshaw06,Spergel06}, we have $\ell_1=220.7\pm0.7$ and finally
\bq
\chi_{\ell_1}^2 = \frac{(\ell_1-220.7)^2}{0.7^2}.
\eq

For the combined SN Ia and CMB data set, the $\chi^2$ is given by
\bq
\chi^2 = \chi^2_{\rm SN_{sel}} + \chi^2_{\ell_1}.
\eq
We employ the Markov Chain Monte Carlo (MCMC) technique to simulate the 
probability distributions of the parameters in the model. 
The following priors on the parameters are adopted: 
$\Omega_{\chi 0}\in(0, 1)$, $\Omega_{\nu 0}\in(0, 1)$,
$\rm log_{10}(\Gamma/10^{-20} s^{-1}) \in(-7.5, 6.5)$ 
and $h_0\in(0.2, 0.9)$. 
We also assume that the energy density of each components is positive, 
$\Omega_{\nu 0}\in(0, 1-\Omega_{\chi 0}-\Omega_{b0}-\Omega_{r0})$ 
is set so that 
$\Omega_{\Lambda}=1-\Omega_{\chi 0}-\Omega_{\nu 0}-\Omega_{b0}-\Omega_{r0} 
\geq 0$. 
We generate six MCMC chains, and each chain contains about 
one hundred thousand points after convergence is reached.
Finally, these chains are thinned to produce about ten thousand points
in the parameter space.
The details of our MCMC code can be found in our earlier paper \cite{Gong07}.

\begin{figure}[htbp]
\includegraphics[scale = 0.35]{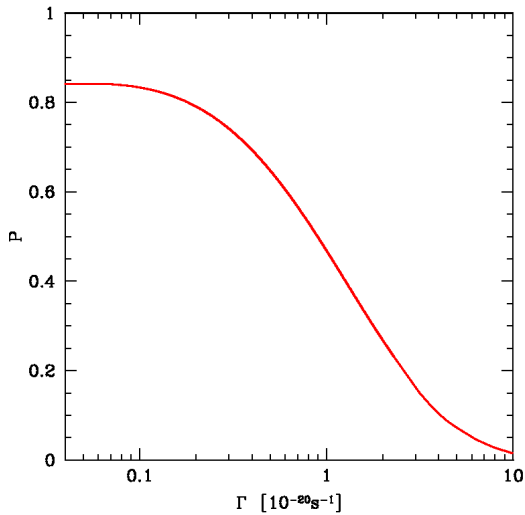} 
\caption{\label{fig:sn_g} The PDF of the dark matter decay-to-neutrino rate
obtained by using the SN Ia and WMAP data.}
\end{figure}

\begin{figure}[htbp]
\includegraphics[scale = 0.35]{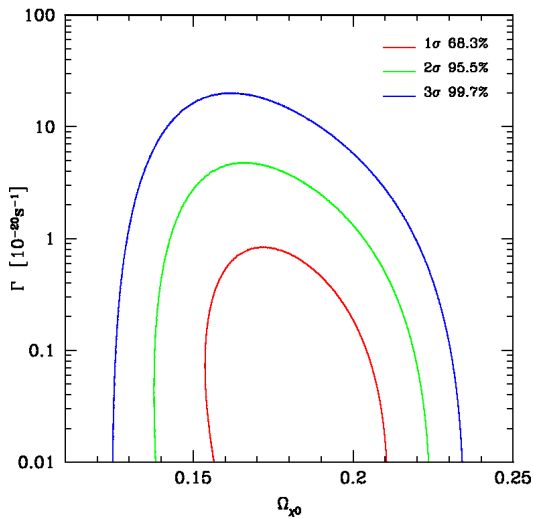} 
\caption{\label{fig:sn_x_g} The contour map of 
$\Omega_{\chi 0}$ vs. $\rm \Gamma$
from SN Ia and CMB data.The $1\sigma(68.3\%)$, 
$2\sigma(95.5\%)$ and $3\sigma(99.7\%)$ 
confidence levels are marked by red, 
green and blue solid lines respectively.}
\end{figure}

The marginalized one-dimensional probability distribution function (PDF)
for the the decay rate $\Gamma$ is shown in 
Fig.~\ref{fig:sn_g}, and the contour map of 
$\Omega_{\chi 0}$ vs. $\rm \Gamma$ is plotted 
in Fig.~\ref{fig:sn_x_g}. We find that the data favors a cold dark 
matter density around $\Omega_{\chi0}=0.186$, and it 
is consistent with a stable (not decaying) 
dark matter. The limit on the decay width $\Gamma$ are
\bq
\Gamma < 0.8 \times10^{-20} s^{-1}, {\rm 68.3\% C.L.}
\eq
and
\bq
\Gamma < 4.5\times10^{-20} s^{-1}, {\rm 95.5\% C.L.}
\eq
respectively. Thus, the decay-to-neutrino rate of the DCDM must satisfy
$\Gamma^{-1}>0.7\times 10^3$ Gyr at 95.5\% C.L. This is 
a much tighter constraint than previous ones \cite{Lattanzi07}. 

\section{Conclusion}

We considered a generic decaying cold dark matter (DCDM) model which 
decays invisibly, producing relativistic neutrinos in the process.  
As the produced neutrinos has an equation of state of 1/3 whereas the 
original cold dark matter has an equation of state of (almost) 0, 
the cosmic expansion rate is affected. 
We constrain the decay-to-neutrino rate $\Gamma$ 
by analyzing the expansion history of the Universe. 
Using a high-quality SN Ia data set selected from 
several resent observations (including Gold06, SNLS and ESSENCE) 
at low and intermediate redshift, and the position of 
the first peak of the CMB angular spectrum 
given by the WMAP three-year data at high redshift, 
we obtain a model-independent constraint on the decay width 
$\Gamma$ by employing the MCMC technique. We find 
$\rm \Gamma<4.5\times10^{-20}s^{-1}$ at 95.5\% C.L. 
i.e., $\Gamma^{-1} > 0.7\times 10^3$ Gyr. One could put further 
constraint on the decay life time of the particle if $\alpha$ is
known. For example, if $\alpha=0.05$, then the decay life time 
$\Gamma_{\chi}^{-1}> 14 $ Gyr, i.e. greater than the current age of the  
Universe.

Recently the terrestrial atmospheric neutrino 
spectra observed by several neutrino detectors such as 
$\rm Fr\acute{e}jus$ \cite{Daum95}, AMANDA 
\cite{Ahrens02,Munich05,Munich07} and 
Super-Kamiokande \cite{Ashie05} has been used to constrain the
annihilation \cite{Beacom06,Yuksel07,PP07} 
or decay \cite{Sergio07} properties of CDM.
It is assumed that the annihilations or decays 
could produce very energetic neutrinos, which would produce a 
peak in the observed atmospheric neutrino background.
These studies generally yields 
tighter constraint on the decay width of the dark matter than obtained
here. For example, for $E_{\nu} \sim 1-10 \GeV$, the atmospheric neutrinos
yields a constraint which is several order of magnitudes stronger than ours.
However, there are certain
limitations on the applicability of such method. 
If the neutrinos produced in the decay is of relatively low energy,
below the threshold energy of the detector, then they would not 
be detected in atmospheric neutrino
experiments. Also, if the decay process
involve multiple particles and produce neutrinos with a spread of energy
distribution, then one may not always be able to identify a sharp peak in the 
atomospheric neutrino experiment data, although in some analyses, e.g. 
that of Ref.\cite{Sergio07}, this effect is not obvious due to the large 
bin size used. Our method is not 
affected by these problems and is therefore a more conservative 
limit which complements their studies.

\begin{acknowledgments}
Our MCMC chain computation was performed on the Supercomputing Center of 
the Chinese Academy of Sciences and the Shanghai Supercomputing
Center. X.C. acknowledges the hospitality of the Moore center of theoretical 
cosmology and physics at Caltech, where part of this research is performed. 
This work is supported by
the National Science Foundation of China under the Distinguished Young
Scholar Grant 10525314, the Key Project Grant 10533010; by the
Chinese Academy of Sciences under grant KJCX3-SYW-N2; and by the 
Ministry of Science and Technology under the National Basic Science
program (project 973) grant 2007CB815401.
\end{acknowledgments}

\newcommand\AL[3]{~Astron. Lett.{\bf ~#1}, #2~ (#3)}
\newcommand\AP[3]{~Astropart. Phys.{\bf ~#1}, #2~ (#3)}
\newcommand\AJ[3]{~Astron. J.{\bf ~#1}, #2~(#3)}
\newcommand\APJ[3]{~Astrophys. J.{\bf ~#1}, #2~ (#3)}
\newcommand\APJL[3]{~Astrophys. J. Lett. {\bf ~#1}, L#2~(#3)}
\newcommand\APJS[3]{~Astrophys. J. Suppl. Ser.{\bf ~#1}, #2~(#3)}
\newcommand\CQG[3]{~Classical Quantum Gravity  {\bf ~#1}, #2~ (#3)}
\newcommand\JCAP[3]{~JCAP. {\bf ~#1}, #2~ (#3)}
\newcommand\LRR[3]{~Living Rev. Relativity. {\bf ~#1}, #2~ (#3)}
\newcommand\MNRAS[3]{~Mon. Not. R. Astron. Soc.{\bf ~#1}, #2~(#3)}
\newcommand\MNRASL[3]{~Mon. Not. R. Astron. Soc.{\bf ~#1}, L#2~(#3)}
\newcommand\NPB[3]{~Nucl. Phys. B{\bf ~#1}, #2~(#3)}
\newcommand\PLB[3]{~Phys. Lett. B{\bf ~#1}, #2~(#3)}
\newcommand\PRL[3]{~Phys. Rev. Lett.{\bf ~#1}, #2~(#3)}
\newcommand\PR[3]{~Phys. Rep.{\bf ~#1}, #2~(#3)}
\newcommand\PRD[3]{~Phys. Rev. D{\bf ~#1}, #2~(#3)}
\newcommand\PPNP[3]{~Prog.Part.Nucl.Phys.{\bf ~#1}, #2~(#3)}
\newcommand\SCIE[3]{~Science {\bf ~#1}, #2~(#3)}
\newcommand\SJNP[3]{~Sov. J. Nucl. Phys.{\bf ~#1}, #2~(#3)}
\newcommand\ZPC[3]{~Z. Phys. C{\bf ~#1}, #2~(#3)}

\end{document}